\documentclass[numbers]{elsarticle}
\usepackage{latexsym}
\usepackage{natbib}
\usepackage[latin2]{inputenc}
\usepackage[hidelinks]{hyperref}
\usepackage{longtable, lineno}

\makeatletter
\def\ps@pprintTitle{%
	\let\@oddhead\@empty
	\let\@evenhead\@empty
	\def\@oddfoot{\centerline{\thepage}}%
	\let\@evenfoot\@oddfoot}
\makeatother

\begin{document}

\title{The Dorsal Striatum and the Dynamics of the Consensus Connectomes in the Frontal Lobe of the Human Brain}


\author[p]{Csaba Kerepesi}
\ead{kerepesi@pitgroup.org}
\author[p]{Bálint Varga}
\ead{balorkany@pitgroup.org}
\author[p]{Balázs Szalkai}
\ead{szalkai@pitgroup.org}
\author[p,u]{Vince Grolmusz\corref{cor1}}
\ead{grolmusz@pitgroup.org}
\cortext[cor1]{Corresponding author}
\address[p]{PIT Bioinformatics Group, Eötvös University, H-1117 Budapest, Hungary}
\address[u]{Uratim Ltd., H-1118 Budapest, Hungary}

\date{}


\begin{abstract}
In the applications of the graph theory it is unusual that one considers numerous, pairwise different graphs on the very same set of vertices. In the case of human braingraphs or connectomes, however, this is the standard situation: the nodes correspond to anatomically identified cerebral regions, and two vertices are connected by an edge if a diffusion MRI-based workflow identifies a fiber of axons, running between the two regions, corresponding to the two vertices. Therefore, if we examine the braingraphs of $n$ subjects, then we have $n$ graphs on the very same, anatomically identified vertex set. It is a natural idea to describe the $k$-frequently appearing edges in these graphs: the edges that are present between the same two vertices in at least $k$ out of the $n$ graphs. 
Based on the NIH-funded large Human Connectome Project's public data release, we have reported the construction of the Budapest Reference Connectome Server \url{http://connectome.pitgroup.org} that generates and visualizes these $k$-frequently appearing edges. We call the graphs of the $k$-frequently appearing edges ``$k$-consensus connectomes'' since an edge could be included only if it is present in at least $k$ graphs out of $n$. Considering the whole human brain, we have reported a surprising property of these consensus connectomes earlier. In the present work we are focusing on the frontal lobe of the brain, and we report here a similarly surprising dynamical property of the consensus connectomes when $k$ is gradually changed from $k=n$ to $k=1$: the connections between the nodes of the frontal lobe are seemingly emanating from those nodes that were connected to sub-cortical structures of the dorsal striatum: the caudate nucleus, and the putamen. We hypothesize that this dynamic behavior copies the axonal fiber development of the frontal lobe.  An animation of the phenomenon is presented at \url{https://youtu.be/wBciB2eW6_8}.
\end{abstract}

\maketitle

\section*{Introduction} 

 The public data releases of the Human Connectome Project (HCP) \cite{McNab2013} make possible to study the high-quality brain imaging data for the scientific community worldwide. Numerous publications apply the data of the HCP as a standard source for studying human cerebral anatomy and also for testing novel methods, e.g., \cite{Barch2013, Calabrese2014,delaIglesia-Vaya2011,Minati2014,Szalkai2016b}. 
 
 One possible application of the HCP data releases is the construction of human anatomical connectomes, describing the macroscopic connections formed by axonal fibers between the cortical and sub-cortical gray matter areas. These graphs have nodes, corresponding to the anatomically identified gray matter areas of 1-1.5 cm$^2$, and two nodes are connected by an edge if a diffusion MRI-based workflow of algorithms finds axonal fibers connecting the areas, corresponding to the two nodes.  
 
 Recently, there is a consideerable interest in the correlations of the  connectome-properties and the biological and psychological characteristics of the subjects \cite{Szalkai2016c,Wang2014,Szalkai2015b,Nakagawa2013,Szalkai2016a,Lynall2010,Szalkai2015c,Diederen2013,Kerepesi2015a}. We believe that even this rough, macroscopic braingraphs with several hundred vertices would lead to deep insights into the functions and the diseases of the human brain.
 
 \subsection*{One vertex set - many graphs:} It is unusual in the applied graph theory that we encounter hundreds of distinct graphs on the very same vertex set: The webgraph \cite{Oerdoeg2014, Brin2012} is a single large graph on billions of vertices; the different interactomes in molecular biology \cite{Ivan2011,Grolmusz2015a,Banky2013,Tothmeresz2013} have distinct vertex sets, as well as the frequently studied large graphs in sociology \cite{Carrington2005} or chemistry \cite{Balasubramanian1985}. 
 
 In the case of braingraphs or connectomes, we generate different edge sets on the same, anatomically identified vertex set for each subject. In other words, we have different graphs for each subject, but the vertex set is the same for all of them. 
 
 This particular scenario makes possible the comparison of the different graphs on the same vertex set in several ways. One possible question is the mapping of the individual variability of the brain connections, measured in distinct anatomical regions. In our work \cite{Kerepesi2015a} we have made those comparisons between the lobes and also between smaller areas of the brain, and we have found different variability in different anatomical regions. 
 
 Another natural question is the description of the common edges in the different graphs. For this goal, we have prepared the Budapest Reference Connectome Server at the address \url{http://connectome.pitgroup.org}. In Version 1.0 from 6 graphs of 5 subjects, in version 2.0 from 96 graphs \cite{Szalkai2015a}, and in version 3.0 from  \cite{Szalkai2016} 418 graphs the consensus connectomes are computed. 
 
 Consensus connectomes are defined as follows. Let us say that an edge appears $k$-frequently if it is present between the same two vertices at least $k$ times out of the all the graphs considered. The $k$-consensus connectome consists of all the $k$-frequent edges. In other words, in a $k$-consensus connectome an edge is present only if there are at least $k$ graphs, where the edge is also present.
 
 The Budapest Reference Connectome Server \url{http://connectome.pitgroup.org} can generate the $k$-consensus connectomes for any $k$ ($1\leq k \leq n$, where $n$ is the number of the graphs processed, i.e., $n=6, n=96, n=418$ or $n=477$, depending on the version), and some other parameters can also be selected \cite{Szalkai2015a,Szalkai2016}. The generated graph can be readily visualized on the webpage and can also be downloaded in CSV and GraphML formats for further processing.
 
 The consensus connectomes are useful in filtering out the rarely appearing connections, therefore, the edges in them have ``strong support'', or many witnesses in the graphs processed. This way the consensus connectomes describe the ``normally'' appearing edges in the connectomes of healthy people (the public release of the Human Connectome Project \cite{McNab2013} contains data from healthy subjects of ages from 22 through 35).

 \begin{figure}[h!]
 	\centering
 	\includegraphics[width=4.6in]{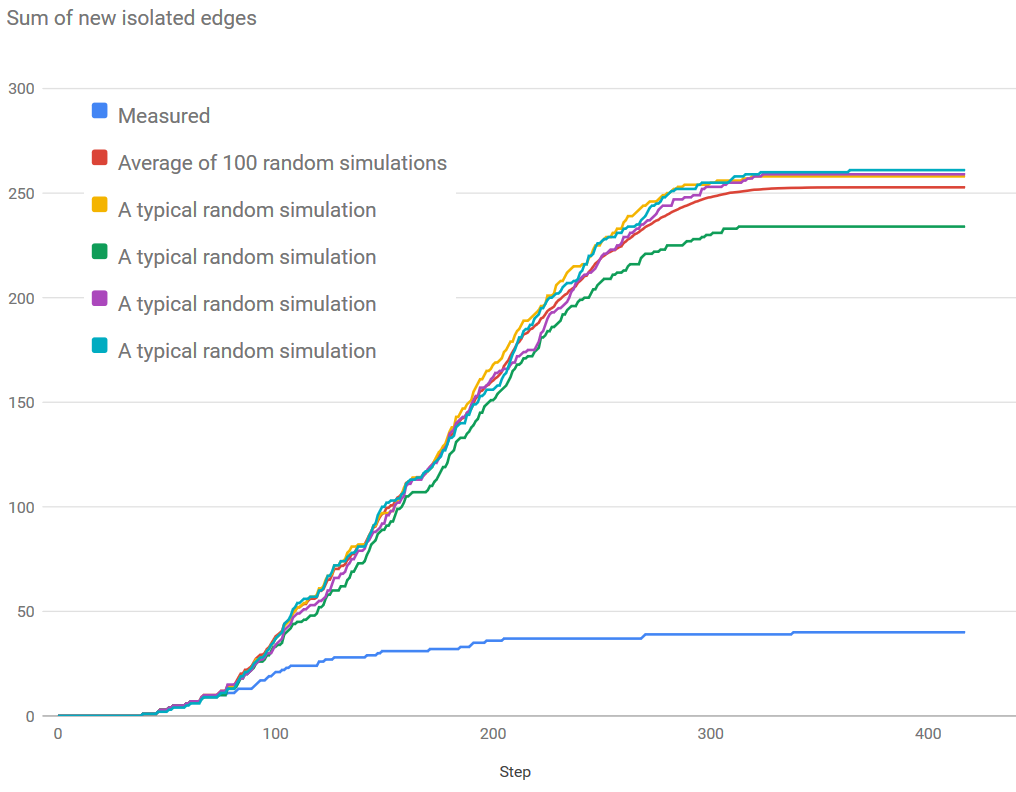}
 	\caption{The comparison of the random simulation and the real buildup of the edges in the $k$-consensus connectomes, restricted to the frontal lobe. Steps are counted from $\ell=0$ to $418$, and for each step $\ell$ we consider the edges that are present in at least 418-$\ell$ graphs. Clearly, for $\ell=0$ we consider the edges that are present in all 418 graphs, then for $\ell=1$ in at least 417 graphs, then in 416, and so on. When $\ell$ increases then the number of edges that are present in at least 418-$\ell$ graphs also increases. The lower, ``Measured'' curve visualizes the sum of the numbers of the newly appearing edges connecting two, previously isolated vertices (we call these edges ``isolated edges'') that are in 418-$\ell$ graphs but were not present in 418-$\ell$+1=419-$\ell$ graphs. The curves of the random simulations show, for each $\ell$, the sum of the new isolated edges that appear when we randomly assign the same number of edges to the pairs of vertices, as the number of new edges appear in the $418-\ell$-consensus connectomes. Clearly, both the number and the sum of the numbers of the isolated edges are much larger in the random simulation then in the case of the measured data in the $k$-consensus connectomes. This observation implies that in the frontal lobe the new edges - typically - are not isolated, that is, they are connecting to the already existing edges when $\ell$ is increasing.}
 \end{figure}
 
 \subsection*{Consensus Connectome Dynamics:} After the publication describing the Budapest Reference Connectome Server \cite{Szalkai2015a} had been appeared, a surprising property of the server was discovered. Let $n$ denote the number of the graphs processed. If we consider $k$-consensus connectomes, decreasing $k=n$ through $k=1$ one-by-one, clearly more and more edges appear in the consensus connectomes. 
 
 We note that anybody can easily experience this on the website \url{http://connectome.pitgroup.org}, after selecting ``Show options'', by moving the ``Minimum edge confidence'' slider from right to left.
 
 The astonishing observation is that the growing number of edges build up a developing, tree-like graph; that is, the edges do not appear randomly, but they appear as the branches of a growing tree. For the whole braingraph we have reported this phenomenon in \cite{Kerepesi2015b} and have visualized that on a video at \url{https://youtu.be/yxlyudPaVUE}.
 
 We call this phenomenon ``Consensus Connectome Dynamics'', and abbreviate it as  ``CCD''.

 \begin{figure} [h!]
 	\centering
 	\includegraphics[width=5in]{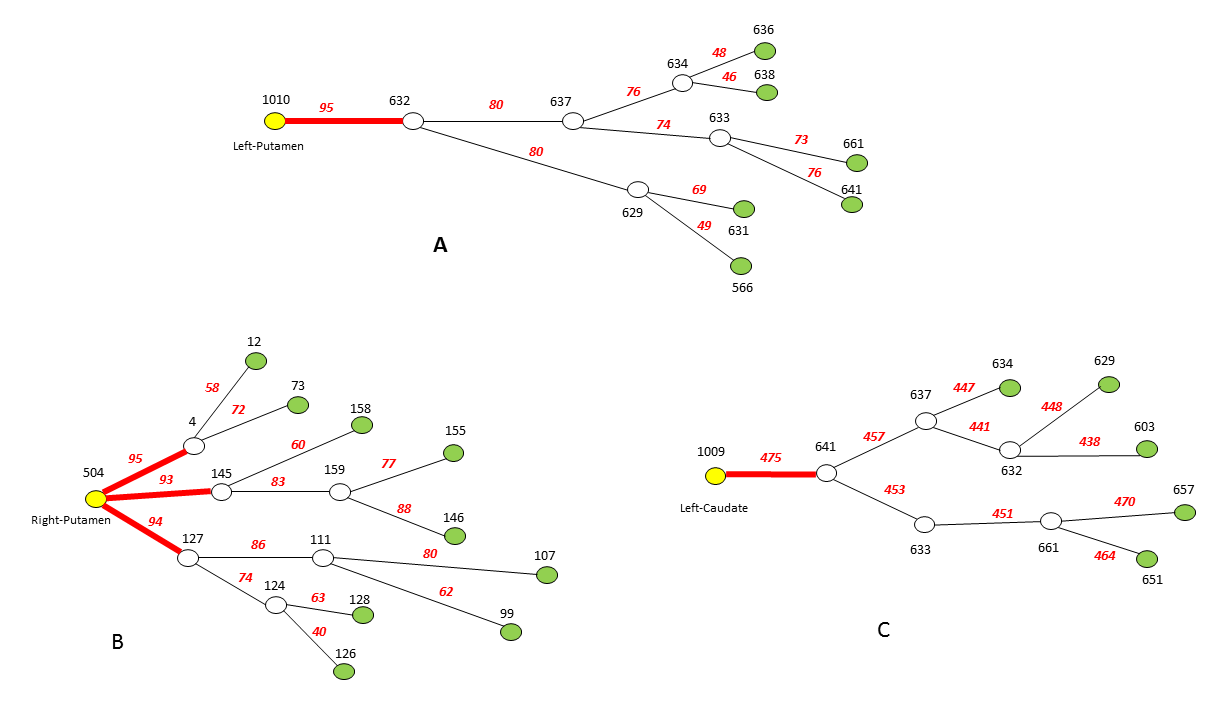}
 	\caption{Three examples of the ``growing subgraphs'' observable in the consensus connectome dynamics in the frontal lobe, originating from the nodes of the dorsal striatum. In all three panels the leftmost yellow node belong to the dorsal striatum, all the other nodes belong to the frontal lobe. The black numbers identify the brain area that corresponds to the node in the 1015-node Desikan-Killiany atlas \cite{Desikan2006} generated by FreeSurfer \cite{Fischl2012}. The red, italic numbers on the edges show the frequency of the edge, that is, the number of graphs in which the edge is present. Note that from left to right the frequencies are mostly (with very few exceptions)  decreasing; that is, in consensus connectome dynamics they are growing from left to right. Panels A and B are made from version 2 and Panel C from version 3 of the Budapest Reference Connectome Server. The red edges show the edges with the largest frequency connecting Left-Putamen and the Left Caudate Nucleus to any vertex to the frontal lobe, on Panel B three edges of the highest frequency were chosen that connect Right-Putamen to nodes of the frontal lobe. Then, from vertex 632 in Panel A we choose the two edge of the highest frequency that connect new vertices (that are not present yet in the structure) to vertex 632. Then we repeat this step for vertices 637, 629, etc. The same building method was applied in Panels B and C. In the case of green vertices we halted the process. Some vertices cannot be connected to two new nodes, just one (e.g., 633 on Panel C). The frequencies of most edges are decreasing from left to right, with some exceptions: e.g., the $\{661,657\}$ edge on Panel C,  or the $\{633,641\}$ edge on Panel A. Node numbers are resolved on the Supplementary Tables S1, S2 and S3. We note that several of the nodes of the frontal lobe are also directly connected to the yellow vertices of the dorsal striatum, but with lower frequencies than the red edges indicated.}
 \end{figure}
 
 \subsection*{The Axonal Development Hypothesis Explains the CCD Phenomenon:} Since the difference from the random appearances of the edges is very clear on the visualization and also in Figure 2 in \cite{Kerepesi2015b}, we have presented a hypothesis in \cite{Kerepesi2015b} as follows: we think that the CCD phenomenon copies the individual development of the cerebral connections in a way that the oldest connections are those that are present in all or almost all connectomes, and gradually the younger connections are appearing as new edges in $k$-consensus connectomes, by decreasing the value of $k$ one-by-one. 
 
 We say that the {\em frequency of an edge} is $k$ if it is present in exactly $k$ of the $n$ graph processed.
 
 Usually, the connections between the sub-cortical structures, especially between the components of the dorsal striatum have the highest frequency, and the next highly frequent connections are between the nodes in the dorsal striatum and the nodes of the cortex,  (see Tables S1, S2 and S3 in the Supporting material). Then the frequency of the cortex-cortex edges decreases, usually the farther the edge from the sub-cortical nuclei, the smaller is the frequency of the edge.
 
 In the present work we are observing and describing the CCD phenomenon in the frontal lobe of the human brain. Our main result is the observation that the tree-like growing structure is also observable in the frontal lobe (see Figures 1 and 2), but there the root-vertices of the trees are those nodes in the frontal lobe that are connected to the  dorsal striatum with the highest frequency edges.

 \section*{Results and Discussion}
 
 The consensus connectome dynamics, restricted to the frontal lobe, is visualized on the video at \url{https://youtu.be/wBciB2eW6_8}. The tree-like structures are clearly visible, but they are not rooted in one or two nodes of the graph. 
 
 On Figure 1 we compared in the frontal lobe the random simulation with the measured appearances of the sum of the number of ``isolated edges'' in the Budapest Reference Connectome Server v3.0. An edge is new in the $k$-consensus connectome if it was not present in the $k+1$-consensus connectome. A new edge is an isolated edge if it connects in  the $k$-consensus connectome two vertices of degree one, which were isolated vertices in the $k+1$-consensus connectome. In the random simulation we have added exactly that many new edge in each step to the graph that appeared in the Budapest Connectome Server, and on the chart we visualized the sum of the isolated edges in each step. Clearly, the random simulation produces much more isolated edges than the $k$-consensus connectomes. This observation shows that in the $k$-consensus connectomes the new edges are much more frequently are connected to old edges than in the random simulation; this fact implies the consensus connectome dynamics also in the frontal lobe. 
 
 Three small examples are presented in Figure 2. It is clearly visible that the red numbers, describing the frequency of the edges on all three panels -- with very few exceptions -- are decreasing from left to right. On the video at \url{https://youtu.be/wBciB2eW6_8} the edges with the highest frequency appear first, then, gradually, the edges with smaller and smaller frequencies. On Figure 2 this temporal sequence would mean the growth of the trees from left to right.
 
  We hypothesize that -- similarly as in the whole braingraph, observed in \cite{Kerepesi2015b} -- the edges whose frequencies are higher were developed in an earlier stage of the brain development than those with lower frequencies. We think that our videos at \url{https://youtu.be/wBciB2eW6_8} for the frontal lobe and \url{https://youtu.be/yxlyudPaVUE} for the whole brain approximately reconstruct this development. As a possible explanation, mentioned also in \cite{Kerepesi2015b}, we think that those neurons that connect to the developing structure will not receive apoptosis signals \cite{Roth2001,Nonomura2013,gordon1995apoptosis} and will survive, while other neurons, which are not connected to the growing graph, will receive apoptosis signals in the individual brain development.

\section*{Methods} 

The program and the methods applied in preparation of the Budapest Reference Connectome Server \url{http://connectome.pitgroup.org} is described in \cite{Szalkai2015a,Szalkai2016}. 

The frontal lobe in our present study comprises the following ROIs of the Desikan-Killiany atlas \cite{Desikan2006}: Superior Frontal, Rostral and Caudal Middle Frontal, Pars Opercularis, Pars Triangularis, Pars Orbitalis, Lateral and Medial Orbitofrontal, Precentral, Paracentral and the 
Frontal Pole. 

The animation at \url{https://youtu.be/wBciB2eW6_8} were prepared by our own Python program from the tables generated by the Budapest Reference Connectome Server \citep{Szalkai2015a,Szalkai2016}. Only those edges are shown that have both endpoints in the frontal lobe. The settings applied: Version 3.0 (i.e., 418 subjects), Population: All (i.e., both male and female subjects), Minimum edge confidence running from 100 \% (418 graphs) through 84\% (353 graphs), Minimum edge weight is 0, Weight calculation model: Median. It contains the common edges found in $k$ subject's braingraphs, from $k=418$ through $k=353$. The number of vertices is 1015.

\section*{Conclusions:} Here we have demonstrated that the astonishing observation of the Consensus Connectome Dynamics, described first in \cite{Kerepesi2015b} for the whole brain, also holds for the frontal lobe. In this case, the roots of the growing structures seem to be those nodes of the frontal lobe, which was connected with the highest frequency to the nodes of the dorsal striatum.

\section*{Data availability:} The pre-processed and the unprocessed MRI data that served as the source of our work are available at the Human Connectome Project's website: 
\url{http://www.humanconnectome.org/documentation/S500} \cite{McNab2013}. 

\noindent The graphs that we assembled from the Human Connectome Project's data and that were subsequently used to build the Budapest Reference Connectome Server can be downloaded at the site \url{http://braingraph.org/download-pit-group-connectomes/}. 

The Budapest Reference Connectome Server is available at \url{http://connectome.pitgroup.org}, there the reader can also generate and download $k$-consensus connectomes. The animation visualizing the consensus connectome dynamics of the frontal lobe is available at \url{https://youtu.be/wBciB2eW6_8}.



\section*{Supporting material}

The supporting Tables S1, S2, S3 are available as a zip file at \url{http://uratim.com/frontal/supporting.zip}.

Table S1 contains the edges that appear in more than 75 from the 96 graphs of the Budapest Reference Connectome Server v2.0, with 1015 nodes. 

Table S2 contains the edges of the Budapest Reference Connectome Server v2.0 with both endpoints in the frontal lobe. From the 1015 nodes, 335 are in the frontal lobe. 

Table S3 contains the edges that appear in more than 367 from the 418 graphs of the Budapest Reference Connectome Server v3.0, with 1015 nodes (weight function: fiber count, minimum edge weight: 70 fibers (that is, only those edges are shown that are defined by at least 70 fibers discovered), weight calculation mode: median, number of fibers launched: 20k).

\section*{Acknowledgments}
Data were provided in part by the Human Connectome Project, WU-Minn Consortium (Principal Investigators: David Van Essen and Kamil Ugurbil; 1U54MH091657) funded by the 16 NIH Institutes and Centers that support the NIH Blueprint for Neuroscience Research; and by the McDonnell Center for Systems Neuroscience at Washington University.

\section*{References}



\end{document}